\documentclass[smallextended,linenumbers]{svjour3}       
\smartqed  
\usepackage{graphicx}
\usepackage{mathptmx}      
\usepackage{natbib} 
\usepackage{hyperref}
\usepackage{amsmath,amssymb,amsfonts}

\begin{document}

\title{Leveraging turbulence data from fusion experiments}

\author{Minjun J. Choi}

\institute{Korea Institute of Fusion Energy \at
              169-148 Gwahak-ro, Yuseong-gu, Daejeon 34133, Republic of Korea \\
              Tel.: +82-42-879-5185\\
              Fax: +82-42-879-6907\\
              \email{mjchoi@kfe.re.kr}
}

\date{Received: date / Accepted: date}

\maketitle

\begin{abstract}
Various methods for leveraging turbulent fluctuation measurements from fusion plasma experiments are introduced, along with selected application examples.   
These can be categorized into spectral methods, statistical methods, and physics informed neural network based methods, and they are most effective for two-dimensional turbulence measurements, which are now widely accessible. 
Extracting more information from turbulence data would pave the way for a better understanding of plasma turbulence transport in fusion experiments. 

\end{abstract}

\section{Introduction}

Plasma density or electron temperature fluctuation is now routinely measured in two-dimensional space in many fusion plasma devices. 
The beam emission spectroscopy (BES)~\citep{ZoletnikPPCF1998, GhimRSI2010, McKee:2010hb, Nam:2014kd} and the microwave imaging reflectometry (MIR)~\citep{Park:2004jd, Lee:2014jw} have been developed to measure the local plasma density fluctuation, and the electron cyclotron emission imaging (ECEI) diagnostics~\citep{Park:2004jd, Yun:2014kv, GaoJINST2018, JiangFED2023, ZhuRSI2020, SabotEPJ2019} for the electron temperature fluctuation.
Analyses of measured data from these diagnostics have revealed the characteristics of low-$k$ plasma turbulence\footnote{Channels of these diagnostics are distributed in the poloidal cross-section of toroidal fusion plasmas, with separations of 1--2~cm in the radial or vertical direction. They can detect fluctuations of $k \rho_s \le 0.3$ in typical plasma conditions where $k$ is the wavenumber and $\rho_s$ is the ion sound Larmor radius.} and advanced the understanding of turbulence transport in fusion experiments. 

In this paper, we provide a brief review of analysis methods for leveraging the measured turbulent fluctuation data for various purposes and under different conditions. 
The linear and nonlinear spectral methods, introduced in section~\ref{sec:spec} can provide the noise suppressed frequency or wavenumber spectrum and detect high order couplings among fluctuation components.  
In particular, the nonlinear spectral methods can identify how the fluctuation energy flows between different modes.  
The statistical perspective and analysis methods, introduced in section~\ref{sec:stat}, would be necessary for dealing with more complex situations or assessing their complexity.
Note that the Python implementation of most spectral and statistical methods is open to public via the GitHub repository:~\url{https://github.com/minjunJchoi/fluctana}. 
In section~\ref{sec:pinn}, the physics informed neural network (PINN)~\citep{RaissiS2020} and its capability with turbulence data are explained, and some practical issues such as the noise and resolution effects are investigated. 
With the help of PINNs, turbulence data can be utilized to predict a missing velocity field in the physics system\footnote{Accurate and high-resolution measurement of velocity fluctuation is very challenging in fusion plasmas, though it is crucial to characterize turbulence and to estimate transport by turbulent fluctuations. Direct measurement is often limited in the particular wavenumber or real space. Indirect approaches have utilized two-dimensional density or temperature fluctuation measurements in time~\citep{EntersRSI2023}, but they are vulnerable to noises and the resolution is limited.} or to directly validate the turbulence model equations. 

\section{Spectral methods}\label{sec:spec}

\subsection{Noise suppressed frequency spectrum}

A frequency spectrum is the most fundamental analysis of time series data, providing an estimate of fluctuation power at each frequency.
In general, the discrete Fourier transform (DFT) is used to estimate the frequency spectrum of discrete time series data~\citep{Welch:1967tq} obtained by diagnostic channels.\footnote{The wavelet~\citep{Torrence:1998jp} or the maximum entropy method~\citep{Haykin1983} based spectrum can be considered when a stationary analysis period is too short or the obtained data is limited. Or, if the data is sampled unevenly, the Lomb-Scargle periodogram~\citep{VanderPlasAJSS2018} can be used to obtain the frequency spectrum.}
Let $X(f_p)$ be the DFT of a discrete time series data $x(t_j)$. 
For each frequency index $p$, the DFT coefficient $X(f_p)=X_p$ is a complex value and it can be written as $X_p = A_p e^{i \alpha_p}$ where $A_p$ represents an amplitude of the $f_p$ frequency sinusoidal oscillation in the time series data and $\alpha_p$ is the initial phase. 
If a mode with the wavenumber $\bold{k}$ and the amplitude $G$ is ideally measured as an oscillation with the frequency $f=f_m$ by some diagnostic channel, we may expect to get $|X_m| = A_m = G$ from the DFT of the recorded signal. 

\begin{figure}
\includegraphics[keepaspectratio,width=1.0\textwidth]{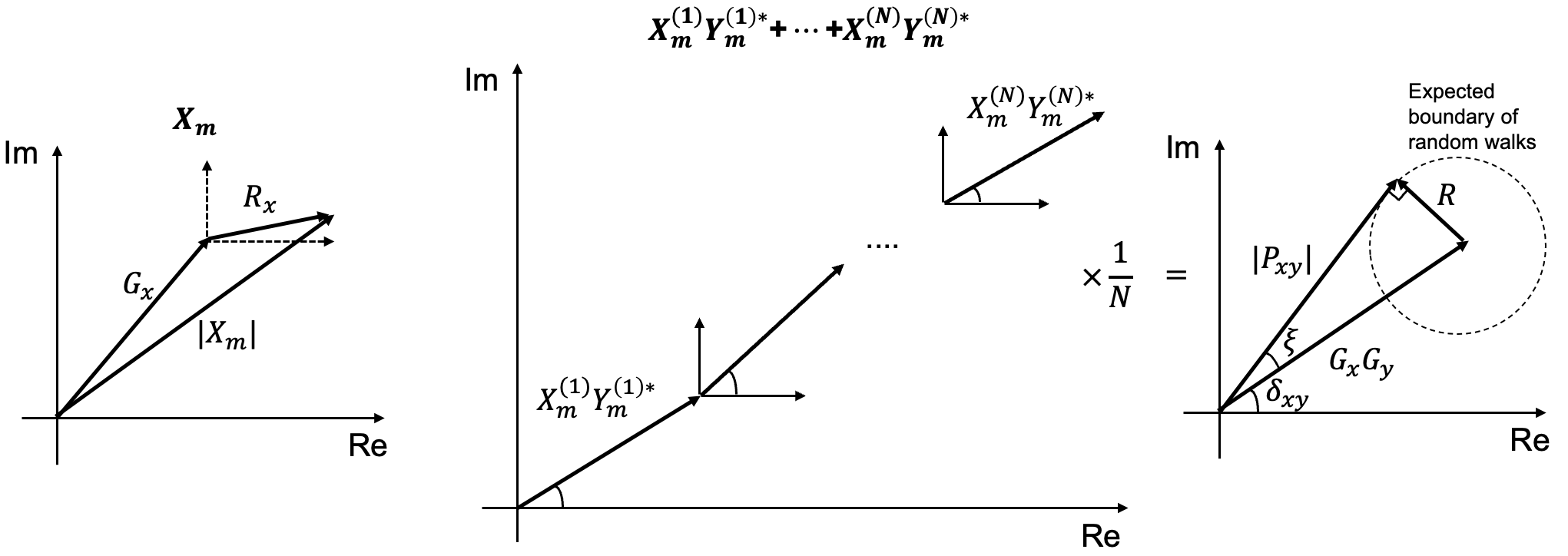}
\centering
\caption{The Fourier transform coefficient in the complex plane and the ensemble average}
\label{fig:complex}
\end{figure}

However, there is always noise component in the measured signal from real experiments. 
$X_m$ will include both the mode component ($G_x e^{i \delta_x}$) and the noise component ($R_x e^{i n_x}$) where $G_x$ and $R_x$ indicate the measured mode and noise amplitudes of the frequency $f_m$ in $x(t_j)$, and $\delta_x$ and $n_x$ are their initial phases, respectively (see figure~\ref{fig:complex}). 
The noise should be suppressed to get the correct information about the mode.

Methods have been developed to suppress noises assuming that they are rapidly decorrelated in time or space~\citep{WattsFST2007}.
Signals from channels separated by a distance would have decorrelated noises which can be reduced by the ensemble average.
If the separation is within the correlation length of a mode, their mode components would be still correlated, meaning that they have a coherent phase relationship surviving the ensemble average (see below).

Two-dimensional diagnostics composed of channels closely distributed in space is ideal to utilize the spatial decorrelation of noise to get the noise suppressed frequency spectrum. 
Let $y(t_j)$ another discrete time series data measured by an adjacent channel separated by $\bold{d}$ and $|\bold{d}| < \lambda_c$ where $\lambda_c$ is the $\bold{d}$-direction correlation length of the mode of wavenumber $\bold{k}$.\footnote{The correlation length of a mode can be estimated as the decay length of the cross correlation of the mode component between a signal at the reference location and signals at other locations. Or, it can be estimated by 1/(the wavenumber spectral width) (see below).}
Let $Y(f_p) = Y_p$ be the DFT of $y(t_j)$. 
At the mode frequency $f_m$, $Y_m$ will include both $G_y e^{i \delta_y}$ and $R_y e^{i n_y}$ where $G_y$ and $R_y$ are the measured mode and noise amplitudes in $y(t_j)$, and $\delta_y$ and $n_y$ are their initial phases, respectively. 
Then, the ensemble average of the $N$ independent measurements of $X Y^*$, the cross power spectrum between signals from two channels, is given as
\begin{equation}
P_{xy} = \langle X Y^* \rangle = \frac{X^{(1)} Y^{(1)*} + ... + X^{(N)} Y^{(N)*}}{N} 
\end{equation}
where the superscripts indicate the measurement number. 
For example, at $f_m$, the 1st measurement can be written as follows.
\begin{align}
X^{(1)}_m Y^{(1)*}_m &= (G_x^{(1)}  e^{i \delta_x^{(1)}} + R_x^{(1)} e^{i n_x^{(1)}})(G_y^{(1)} e^{-i \delta_y^{(1)}} + R_y^{(1)} e^{-i n_y^{(1)}})  \nonumber \\
&= G_x^{(1)}G_y^{(1)}  e^{i (\delta_x^{(1)}-\delta_y^{(1)})} + G_x^{(1)} R_y^{(1)}  e^{i (\delta_x^{(1)}-n_y^{(1)})} \nonumber \\
&~+ R_x^{(1)} G_y^{(1)} e^{i (n_x^{(1)}-\delta_y^{(1)})} + R_x^{(1)} R_y^{(1)} e^{i (n_x^{(1)}-n_y^{(1)})}
\end{align}
where each term can be thought as a vector in the complex plane.
Adding up vectors in the complex plane with the random phase ($\delta_x - n_y$, $n_x - \delta_y$, and $n_x - n_y$) will be like a random walk (RW) diffusion whose expected deviation is $\mathrm{(step~size)}\times \sqrt{N}$.
The division by $N$ in the ensemble average will make the random phase terms decay with $1/\sqrt{N}$. 
However, the first term has the non-random coherent phase, $\delta_x^{(1)}-\delta_y^{(1)}$, which is the definite phase difference in mode components of two separated channels, i.e. $\delta_x^{(1)}-\delta_y^{(1)} \equiv \delta_{xy} = \bold{k} \cdot \bold{d}$. 
The ensemble average will lead to  
\begin{equation}
\langle X_m Y_m^* \rangle = G_x G_y e^{-i \delta_{xy}} + \frac{\mathrm{RW}(G_x R_y) + \mathrm{RW}(G_y R_x) + \mathrm{RW}(R_x R_y)}{N}
\end{equation}
where $\mathrm{RW}(C)$ represents the random walk with the amplitude $C$ in the complex plane (see figure~\ref{fig:complex}). 
The amplitude of the cross power can be written as 
\begin{equation}
|\langle X_m Y_m^* \rangle| \approx G_x G_y \pm \mathcal{O}\left(\frac{G_x R_y}{\sqrt{N}}\right) \pm \mathcal{O}\left(\frac{G_y R_x}{\sqrt{N}}\right) \pm \mathcal{O}\left(\frac{R_x R_y}{\sqrt{N}}\right) 
\end{equation}
The ensemble average in the cross power enables a significant noise reduction for the large $N$, providing the noise suppressed frequency spectrum. 

It is convenient to normalize the cross power spectrum with the auto power spectra as follows.
\begin{equation}
\gamma_{xy} = \frac{|\langle X Y^* \rangle|}{\sqrt{\langle XX^* \rangle} \sqrt{\langle YY^* \rangle}}
\end{equation}
It is called the coherence and ranges from 0 to 1, meaning the coherent ($\bold{k} \cdot \bold{d} \approx \mathrm{const}$) fluctuation power fraction against the total power at each frequency.
The noise floor in the coherence will be given as $\frac{|\langle R_x R_y e^{-i(n_x - n_y)} \rangle|}{R_x R_y} \approx 1/\sqrt{N}$. 

\subsubsection{Practical example}

\begin{figure}
\includegraphics[keepaspectratio,width=0.5\textwidth]{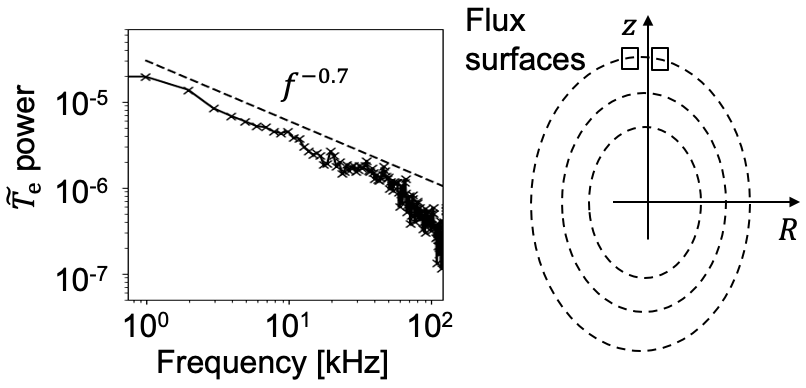}
\centering
\caption{The power-law power spectrum indicates that the event size would exhibit the power-law distribution. Two adjacent channels (square boxes) on the same flux surface are used to reduce noise components in the power spectrum. Reprinted from~\citep{ChoiNF2019}. Copyright 2019 IOP}
\label{fig:avalanche}
\end{figure}

An accurate measurement of the frequency spectrum is not only fundamental but also provides valuable insights into plasma transport. 
The non-diffusive transport model based on the self-organized criticality suggests that ballistic transport events, called avalanches, of the power-law size spectrum dominate plasma transport~\citep{Diamond:1995hz, Hahm:2018dm}. 
In the KSTAR plasma in which the magnetohydrodynamic instabilities are quiescent, the ballistic transport events are observed to limit the electron energy confinement~\citep{ChoiNF2019, ChoiPPCF2024}.
They appear as electron temperature bumps ($\delta T_\mathrm{e} > 0$) and voids ($\delta T_\mathrm{e} <0$) which are radially propagating outwards and inwards, respectively. 
Two ECEI channels on the same flux surface were able to measure the noise suppressed cross power spectrum of the $\delta T_\mathrm{e}$ size of the events. 
Figure~\ref{fig:avalanche} shows the result exhibiting the power-law behavior $S(f) \propto f^{-0.7}$.
The accurate cross power frequency spectrum could identify the avalanche characteristics of the electron heat transport events in this plasma~\citep{ChoiNF2019}. 

\subsubsection{Frequency spectrum with a long time correlated noise}

\begin{figure}
\includegraphics[keepaspectratio,width=0.45\textwidth]{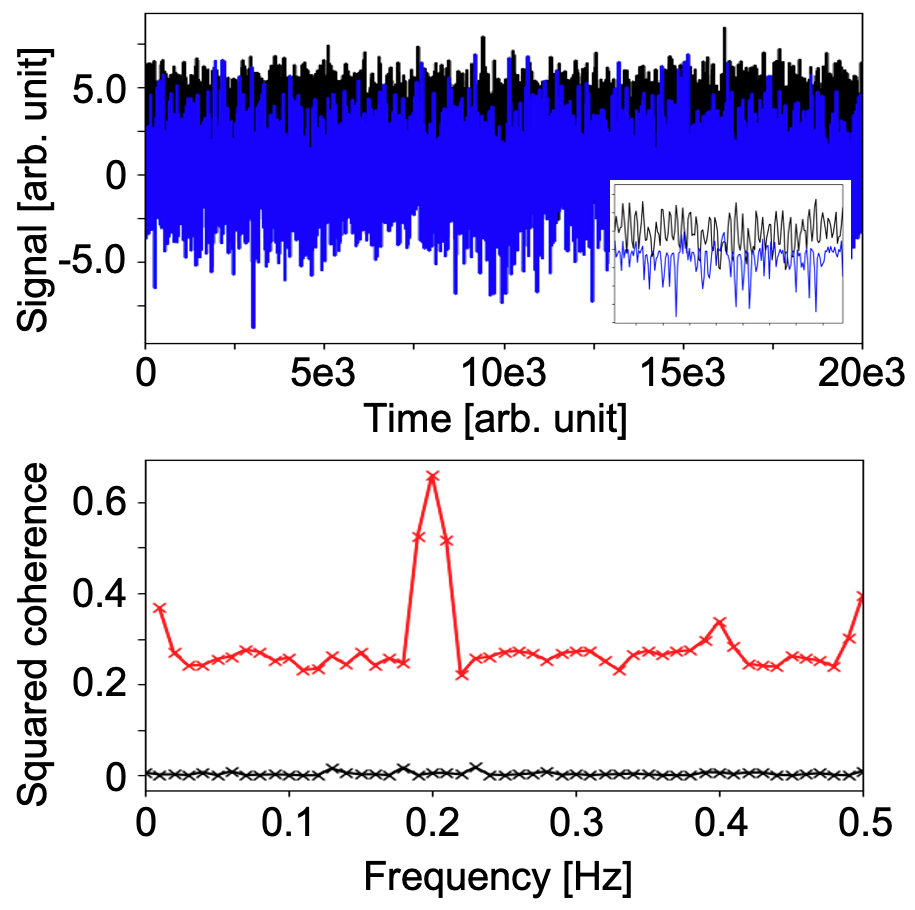}
\centering
\caption{(Color online) (top) Two time series data: $x(t_j) = 2 a_j + \varepsilon_{1j}$ (black) and $y(t_j) = a_j \varepsilon_{2j}$ (blue) where $a_j$ is a sinusoidal signal and $\varepsilon_{1,2}$ are FARIMA(0, 0.3, 0) processes. (bottom) The squared ordinary coherence (black crosses) and squared Laplace coherence (red crosses). }
\label{fig:quantile}
\end{figure}

So far, we have assumed that a signal consists of two additive components, namely a mode component and a noise component.
The noise component is assumed to have a very short correlation time or length like the Gaussian white noise, and it could be suppressed by averaging out independent measurements.
We can consider a more complicated situation where a multiplicative component with the long correlation time is involved.\footnote{A signal of fusion plasma turbulence diagnostics often depends on the product of multiple plasma variables such as density and temperature.} 
For example~\citep{LimAEL2021}, $x(t_j) = 2 a_j + \varepsilon_{1j}$ and $y(t_j) = a_j \varepsilon_{2j}$ where $a_j = 1.0 + 0.9 \cos(\omega_0 t_j)$ is a periodic signal with $\omega_0 = 2 \pi \times 0.2$, and $\varepsilon_{1,2}$ are independent fractionally autoregressive integrated moving average processes, FARIMA(0, 0.3, 0), with the Gaussian distribution. 
Figure~\ref{fig:quantile} shows $x(t_j)$ (black) and $y(t_j)$ (blue) and their squared cross coherence spectra obtained by the ordinary DFT based method described in previous section (black crosses) and by the Laplace cross periodogram~\citep{LiESPC2010}.
While the ordinary coherence could not clearly identify the common periodic signal with frequency of 0.2 in $x(t_j)$ and $y(t_j)$, it could be identified in the Laplace coherence which is more robust against the outliers~\citep{LiESPC2010}. 

\subsection{The local dispersion relation}

The phase difference between the measured signal $x(t_j)$ and $y(t_j)$ from two channels separated by $\bold{d}$ can be calculated as
\begin{equation}
\theta_{xy} = \tan^{-1}\left( \frac{\mathrm{Im}[P_{xy}]}{\mathrm{Re}[P_{xy}]} \right)
\end{equation}
where $P_{xy}$ is the cross power.
It can be written as $\theta_{xy} = \delta_{xy} \pm \mathcal{O}(\xi)$ where $\delta_{xy}$ is the phase difference between the mode components in two signals ($\delta_{xy} = \bold{k} \cdot \bold{d}$) and $\mathcal{O}(\xi)$ is the noise contribution (see figure~\ref{fig:complex}).
Since the maximum possible noise contribution is  
\begin{equation}
\xi = \sin^{-1} \left[ \frac{R}{G_x G_y} \right] \approx \frac{2 \epsilon}{\sqrt{N}} + \frac{\epsilon^2}{\sqrt{N}} + \mathcal{O}\left(\frac{\epsilon^3}{N^{3/2}}\right ) + ...
\end{equation}
where $R = \frac{G_x R_y}{\sqrt{N}} + \frac{G_y R_x}{\sqrt{N}} + \frac{R_x R_y}{\sqrt{N}} $ and $\epsilon = |R_x/G_x| \approx |R_y/G_y| < 1 $, for the large number of ensemble $N$ or the large signal-to-noise ratio $1/\epsilon \gg 1$, $\theta_{xy} \approx \delta_{xy} = \bold{k} \cdot \bold{d}$. 
It means that we can estimate the local wavenumber in the direction of $\bold{d}$ by $\theta_{xy}/d$.

The cross phase frequency spectrum $\theta_{xy}(f)$ would correspond to the local wavenumber frequency spectrum, or the local dispersion relation $K(\omega) = \delta_{xy}(\omega)/d \approx \theta_{xy}(\omega)/d$ where $K$ is the local wavenumber along the $\bold{d}$ direction and $\omega = 2 \pi f$ is the measured angular frequency. 
By using channels of two-dimensional diagnostics separated in the poloidal or radial direction within the correlation length of modes, the local poloidal or radial dispersion relation for the modes can be obtained, respectively.

When the local dispersion relation is found, the phase velocity ($\omega / K$) or the group velocity ($\partial \omega / \partial K$) of a mode in the laboratory frame can be estimated.
Since $\omega$ includes the Doppler shift by the plasma flow ($\omega_D = K v$ where $v$ is the plasma flow along the $\bold{d}$-direction), the laboratory frame velocity measurements ($v_L$) contain some information about the plasma flow, i.e. $v_L = v_P + v$ where $v_P$ represent the plasma frame velocity measurements. 
If we can assume that $v_P$ is nearly uniform in the region of interest, the spatial variation of $v_L$ would come from the spatial variation of the plasma flow. 

\subsubsection{Practical example}

\begin{figure}
\includegraphics[keepaspectratio,width=0.45\textwidth]{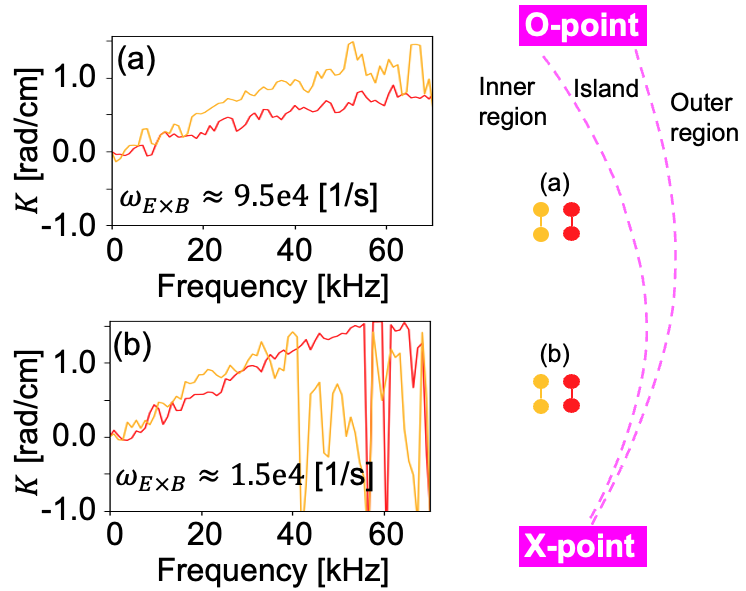}
\centering
\caption{(Color online) The local dispersion measurements at different locations to estimate the flow shearing rate outside a magnetic island. The estimated shearing rate is larger near (a) the O-point than (b) the X-point, consistent with the weaker fluctuation power near the O-point than the X-point~\citep{Choi:2017ez}. Reprinted from~\citep{Choi:2021fs}. Copyright 2021 Springer Nature}
\label{fig:flow}
\end{figure}

The radial shear of the plasma flow is known to suppress an instability whose growth rate is smaller than the flow shear~\citep{Hahm:1995eb}. 
Figure~\ref{fig:flow} shows four local dispersion measurements at different locations in the inner region, near a magnetic island in a KSTAR plasma.
Each local dispersion is obtained from the cross phase frequency spectrum ($\theta_{xy}(f)/d \approx K(f)$) between poloidally adjacent ECEI channels at the same radial location, providing an estimate of the laboratory frame poloidal phase velocity. 
Dispersion measurements at two radial locations (orange and red) near near the O-point (X-point) are shown in figure~\ref{fig:flow}(a) (figure~\ref{fig:flow}(b)), and the O-point measurements indicate the stronger radial shear of the poloidal plasma flow (see figure 5 of the reference~\citep{Choi:2017ez} for the full 2D measurement).
The increasing flow shear towards the O-point could explain that the measured fluctuation power is reduced below the noise level near the O-point~\citep{Choi:2017ez}. 
This radial shear of the flow around the island can be attributed to the $E \times B$ flow perturbation by the magnetic island~\citep{Hornsby:2010fh, Kwon:2018bi}. 

\subsection{Noise suppressed local wavenumber-frequency spectrum}

The local wavenumber-frequency spectrum $S_L(K,f)$ can be estimated using the two-point method introduced in the reference~\citep{Beall:1998fx} as follows
\begin{equation}
S_L(K,f) = \left \langle \left( \frac{X(f)X^*(f) + Y(f)Y^*(f)}{2} \right) \delta_D \left( \frac{\theta_{XY}(f)}{d} - K \right) \right \rangle
\label{Elfw}
\end{equation}
where $\delta_D(\cdot)$ represents the Dirac delta function and $\theta_{XY} = \tan^{-1}\left( \frac{\mathrm{Im}[X Y^*]}{\mathrm{Re}[X Y^*]} \right)$ is a cross phase between each pair of $X$ and $Y$ measurements. 
This provides how the fluctuation power is distributed in the wavenumber $K$ and frequency $f$ space by means of a histogram.
Compared to the local dispersion relation, $S_L(K,f)$ can provide additional information such as a spectral width.
For example, the wavenumber spectral width can be estimated from the second moment in wavenumber, $\sigma_K(f) = \sqrt{\sum_K (K - \overline{K}(f))^2 S_L(K,f)}$ where $\overline{K}(f) = \sum_K K S_L(K,f)$ is the first moment. 
This $\sigma_K(f)$ provides an estimate of the correlation length ($1/\sigma_K(f)$)~\citep{Beall:1998fx, Xu:2006dg}. 

Two-dimensional diagnostics would allow a noise suppressed estimation of $S_L(K,f)$, since multiple pairs of the channels along the same direction can be utilized to perform the ensemble average $\langle S_L(K,f) \rangle_p$ where $\langle \cdot \rangle_p$ represents the average over pairs~\citep{ChoiPoP2022}.\footnote{In fusion experiments, most two-dimensional diagnostic can cover the only limited area. Therefore, the two-point method or the maximum entropy method has been utilized to obtain the wavenumber spectrum with the limited spatial channels~\citep{Kobayashi:2021go}.}

\subsubsection{Practical example}

\begin{figure}
\includegraphics[keepaspectratio,width=0.45\textwidth]{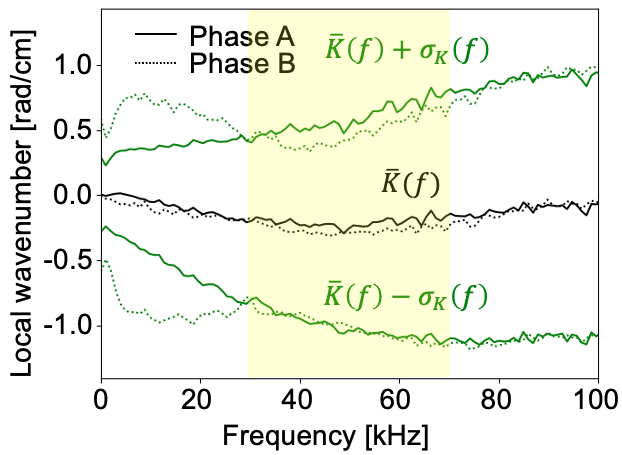}
\centering
\caption{(Color online) The estimated local wavenumber $\overline{K}(f) = \sum_K K \langle S_L(K,f) \rangle_p$ and the spectral width $\sigma_K(f) = \sqrt{\sum_K (K - \overline{K}(f))^2 \langle S_L(K,f) \rangle_p}$ in two different edge localized mode suppression phases A and B by the resonant magnetic perturbation field. Reprinted from~\citep{ChoiPoP2022}. Copyright 2022 AIP}
\label{fig:clength}
\end{figure}

Figure~\ref{fig:clength} shows $\overline{K}(f)$ (black) and $\overline{K}(f) \pm \sigma_K(f)$ (green) from $\langle S_L(K,f) \rangle_p$ measurements using electron temperature fluctuations near the pedestal top in two distinguished edge localized mode (ELM) suppression phases (A and B) by the resonant magnetic perturbation field in KSTAR~\citep{ChoiPoP2022}. 
The poloidal wavenumber spectral width, or 1/(the poloidal correlation length), in the highlighted frequency range (30--70 kHz) is found to be bigger in the phase A than in phase B, meaning the smaller correlation length in the phase A. 
This can be attributed to the stronger field penetration~\citep{Xu:2006dg} in the pedestal top region in the phase A, consistent with the more stochastic behavior of temperature fluctuation in that phase~\citep{ChoiPoP2022}. 

\subsection{Bicoherence to detect three-wave coupling}

The nonlinear three-wave coupling can be identified using the fact that the coupled modes have a coherent phase relation~\citep{Kim:1979ps}. 
Consider three modes with the amplitudes $G_1$, $G_2$, and $G_3$ and frequencies $f_1$, $f_2$, and $f_3=f_1+f_2$ (satisfying the frequency resonance condition) measured by a single channel.
The auto bispectrum is defined as follows
\begin{equation}
B(f_1, f_2) = \langle X_1 X_2 X^*_3 \rangle 
\end{equation}
where $X_{1,2,3}$ indicates the DFT coefficients of $x(t_j)$ at frequencies $f_{1,2,3}$, respectively. 
The 1st measurement in the ensemble average will be 
\begin{equation}
X_1^{(1)} X_2^{(1)} X_3^{(1)*} = G_1^{(1)} G_2^{(1)} G_3^{(1)} e^{i(\delta_1^{(1)} + \delta_2^{(1)} - \delta_3^{(1)})} + ...
\label{eq:bispec}
\end{equation}
where $\delta_1$, $\delta_2$, and $\delta_3$ represent the initial phases of modes. 
If three modes appear independently, their phase difference ($\delta_1 + \delta_2 - \delta_3$) would be more or less random for each measurement or realization.
Then, the $G_1 G_2 G_3$ term would decay with $1/\sqrt{N}$ by the ensemble average. 
However, if they are coupled, $\delta_1 + \delta_2 - \delta_3 = \mathrm{const}$ and the $G_1 G_2 G_3$ term would not decay by the average. 

In practice, the squared bicoherence (the normalized squared bispectrum) is often used to measure the degree of the nonlinear coupling, because we have finite $N$ and the $1/\sqrt{N}$ decay may not be sufficient for large uncorrelated amplitudes to decay in the bispectrum.
The squared bicoherence is defined as follows
\begin{equation}
b^2(f_1, f_2) = \frac{|\langle X_1 X_2 X^*_3 \rangle |^2}{\langle |X_1X_2|^2 \rangle \langle |X_3|^2 \rangle}
\end{equation}
This value ranges from 0 to 1, meaning the fraction of the coupled power at $f_3$ against the total power at $f_3$. 
$\langle |X_3|^2 \rangle$ represents the total power at $f_3$ and $\frac{|\langle X_1 X_2 X^*_3 \rangle |^2 }{\langle |X_1X_2|^2 \rangle}$ is the power at $f_3$ associated with the three-wave coupling with $f_1$ and $f_2$~\citep{Kim:1979ps}. 

To detect the three-wave coupling, the auto bicoherence rather than the cross bicoherence using two or more signals would be preferred because it would not need to care about conditions like $\bold{d} < \lambda_c$. 
The noise components at different frequencies are normally independent and the terms involving the noise component in equation~\eqref{eq:bispec} would decay by the ensemble average.
For the cross bispectrum or bicoherence using multiple signals from the spatially separated channels, in addition to the $\bold{d} < \lambda_c$ condition, $\bold{d} \cdot \nabla v \sim 0$ should be also satisfied since the different Doppler shifts at different channel locations by the flow shear would yield different measured frequency for the same mode.  
To ensure the frequency resonance condition to hold as expected in the calculation $f_3 = f_1 + f_2$, the flow shear in the channel separation direction should be negligible. 
Nonetheless, the cross bispectrum can be useful to investigate the nonlinear energy transfer as described in the next subsection. 


\subsection{Quadratic energy transfer estimation}

Consider a simple system in which the third mode ($f_3$) is purely driven by a sole three-wave coupling with the other two modes, i.e. $G_3 = Q_{1,2}^{3} G_1 G_2$ where $Q_{1,2}^3$ is the coupling coefficient. 
The coupling coefficient can be directly estimated using the bispectrum as $Q_{1,2}^{3} \approx \frac{B^*(f_1,f_2)}{\langle | X_1 X_2 |^2 \rangle}$ for the large signal-to-noise ratio~\citep{Kim:1979ps}.
However, a real system is more complicated and a mode evolves through more than one nonlinear coupling.

For some physical quantity $\phi (z,t)$ at position $z$ and time $t$, we can decompose $\phi (z,t)$ into the summation of the wavenumber Fourier coefficients such as $\phi (z,t) = \sum_p \Phi (k_p, t) e^{i k_p z}$ and study the temporal evolution of each $k_p$ component~\citep{Ritz:1989ci}, or decompose it into the summation of the frequency Fourier coefficients such as $\phi (z,t) = \sum_p \Phi (z,\omega_p) e^{i \omega_p t}$ and study the spatial evolution of each $\omega_p$ component~\citep{Wit:1999jm}. 
For example, the spatial evolution of a mode $\omega_p$ including many quadratic couplings can be written as 
\begin{equation}
\frac{\partial \Phi(z, \omega_p) }{\partial z} = \Lambda_{\omega_p}^L \Phi (z, \omega_p) +  \displaystyle\sum_{\substack{p_1 \ge p_2 \\ p=p_1+p_2}}  \Lambda_{\omega_{p_1},\omega_{p_2}}^Q \Phi (z, \omega_{p_1}) \Phi (z, \omega_{p_2})
\label{eq:quad}
\end{equation}
 where $\Lambda_{\omega_p}^L$ and $\Lambda_{\omega_{p_1},\omega_{p_2}}^Q$ are the linear and quadratic coupling coefficients, respectively. 
We can discretize equation~\eqref{eq:quad} to reconstruct the following system to utilize the two-point measurements at $z$ and $z+d$~\citep{Ritz:1986gh}. 
\begin{equation}
Y_p = L_p X_p + \sum_{\substack{p_1 \ge p_2 \\ p=p_1+p_2}} Q_{p_1,p_2}^{p} X_{p_1} X_{p_2} 
\end{equation}
where $X_p$ and $Y_p$ represent the measured Fourier coefficients at $z$ and $z+d$, respectively, $L_p$ is the linear transfer function, and $Q_{p_1,p_2}^{p}$ is the quadratic transfer function.
Our task is to estimate those transfer functions which transform $X_p$ into $Y_p$. 

The linear and quadratic transfer functions for such a system can be written as follows with the Millionshchikov hypothesis ($\langle X_{p_1} X_{p_2} X_{p_3}^* X_{p_4}^* \rangle \approx \langle |X_{p_1} X_{p_2}|^2 \rangle$)~\citep{Ritz:1986gh,Kim:1996jv}.
\begin{align}
L_p &= \frac{ \langle Y_p X^*_p\rangle - \displaystyle\sum_{\substack{p_1 \ge p_2 \\ p=p_1+p_2}} \frac{\langle X_{p_1} X_{p_2} X^*_p \rangle \langle X^*_{p_1} X^*_{p_2} Y_p \rangle}{\langle | X_{p_1} X_{p_2} | ^2 \rangle} }{ \langle X_p X^*_p \rangle - \displaystyle\sum_{\substack{p_1 \ge p_2 \\ p=p_1+p_2}} \frac{| \langle X_{p_1} X_{p_2}X^*_p \rangle|^2}{\langle |X_{p_1} X_{p_2}|^2 \rangle} } \\
Q_{p_1,p_2}^{p} &= \frac{\langle X^*_{p_1} X^*_{p_2} Y_p \rangle - L_p \langle X^*_{p_1} X^*_{p_2} X_p \rangle}{\langle | X_{p_1} X_{p_2} |^2 \rangle}
\end{align}
Or, we can directly solve the following matrix equation for $\bold{H}$ with the large number of ensembles ($N \gg$ the number of unknowns in $\bold{H}$)~\citep{Wit:1999jm}.

\begin{figure}
\includegraphics[keepaspectratio,width=1.0\textwidth]{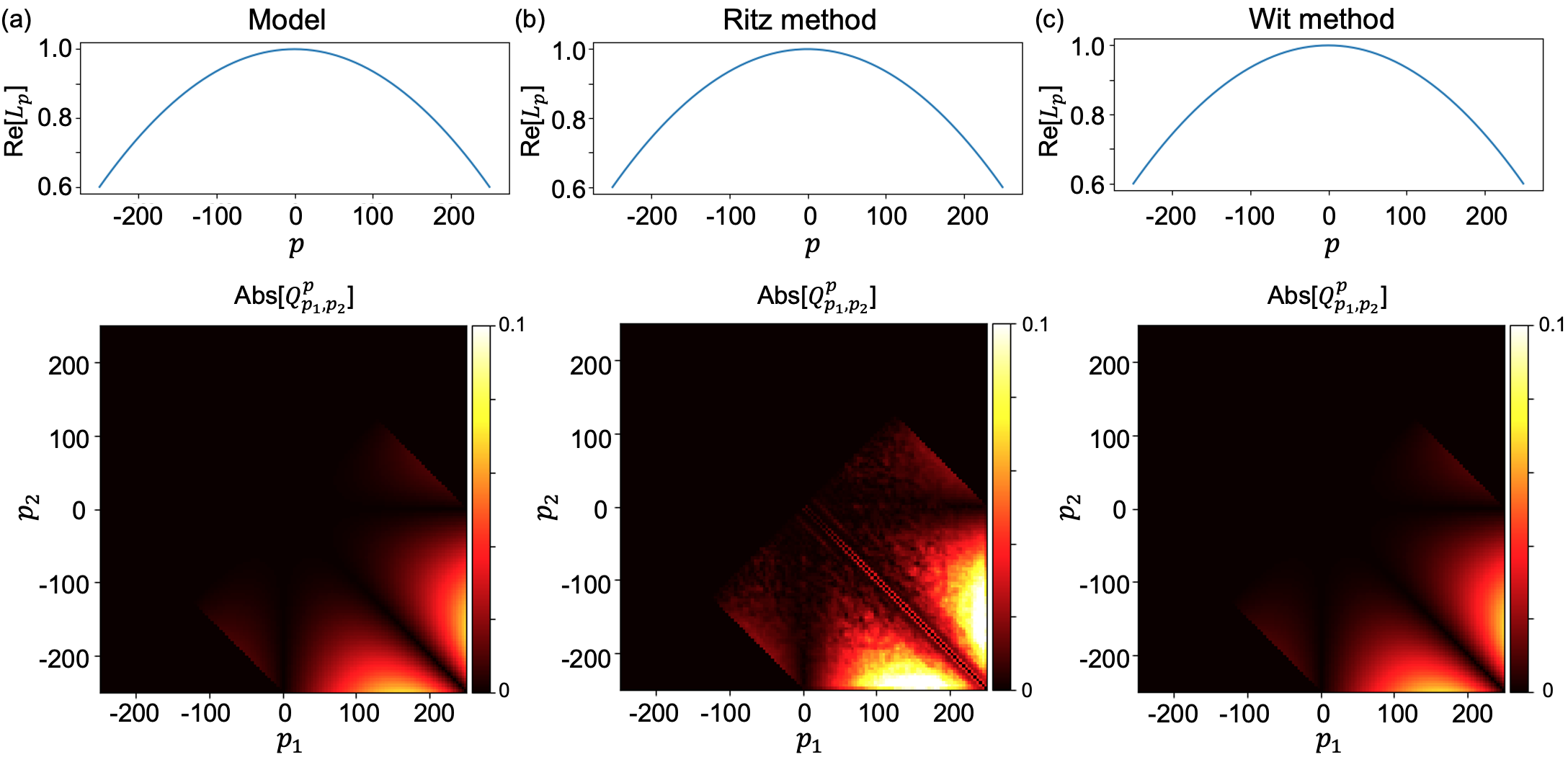}
\centering
\caption{(Color online) (a) The real part of $L_p$ and $|Q_{p_1,p_2}^p|$ from the model equation. Estimation results using (b) the Ritz and (c) Wit method.}
\label{fig:ritzwit}
\end{figure}

\begin{equation}
\bold{Y} = \bold{X} \bold{H}
\end{equation}
where
\begin{align}
\bold{Y} = 
	\begin{bmatrix}
	Y_p^{(1)} \\
	Y_p^{(2)} \\
	\vdots \\
	Y_p^{(N)} \\
	\end{bmatrix}
~\bold{X} = 
	\begin{bmatrix}
	X_p^{(1)}  & X_{p_1}^{(1)} X_{p-p_1}^{(1)} & X_{p_2}^{(1)} X_{p-p_2}^{(1)} & ... \\
	X_p^{(2)}  & X_{p_1}^{(2)} X_{p-p_1}^{(2)} & X_{p_2}^{(2)} X_{p-p_2}^{(2)} & ... \\
	\vdots & \vdots & \vdots & ... \\
	X_p^{(N)} & X_{p_1}^{(N)} X_{p-p_1}^{(N)} & X_{p_2}^{(N)} X_{p-p_2}^{(N)} & ... \\
	\end{bmatrix}
~\bold{H} = 
	\begin{bmatrix}
	L_p \\
	Q_{p_1,p-p_1}^p\\
	Q_{p_2,p-p_2}^p\\
	\vdots \\
	\end{bmatrix}
\end{align}
Verification of these two methods (the Ritz~\citep{Ritz:1986gh,Kim:1996jv} and Wit~\citep{Wit:1999jm} methods) using the model $L_p$ and $Q_{p_1,p_2}^p$ are shown in figure~\ref{fig:ritzwit}. 
The following model equations are used for $L_p$ and $Q_{p_1,p_2}^p$ to generate $Y_p$ from some given $X_p$ data~\citep{Ritz:1986gh}.
\begin{align}
L_p &= 1.0 - 0.4 \frac{p^2}{p_{Nyq}^2} + i0.8 \frac{p}{p_{Nyq}} \\
Q_{p_1,p_2}^p &= \frac{i}{5 p_{Nyq}^4} \frac{p_1 p_2 (p_2^2 - p_1^2)}{1 + p^2/p_{Nyq}^2}
\end{align}
where $p = p_1 + p_2$ and $p_{Nyq}$ is the index of the Nyquist frequency (250 in this case). 

%

Once the transfer functions ($L_p$ and $Q_{p_1,p_2}^p$) are estimated with $X_p$ and $Y_p$ either using the Ritz (with the Millionshchikov hypothesis) or Wit method, we can calculate a spatial linear growth rate $\gamma_p$ and a spatial quadratic energy transfer rate $T_p$ of the spectral power $P_p = \langle X_p X_p^* \rangle$ whose evolution is described as
\begin{equation}
\frac{\partial P_p}{\partial z} \approx \frac{\langle Y_p Y_p^* \rangle - \langle X_p X_p^* \rangle}{d} = \gamma_p P_p + T_p
\end{equation}
where~\citep{Kim:1996jv}
\begin{align}
\gamma_p &\approx \frac{|L_p|^2 - 1}{d} \\
T_p &\approx 2~\mathrm{Re} \left[ L_p^* \sum_{\substack{p_1 \ge p_2 \\ p=p_1+p_2}} \frac{Q_{p1,p2}^p \langle X_{p1} X_{p2} X_p^* \rangle}{d}   \right] + \sum_{\substack{p_1 \ge p_2 \\ p=p_1+p_2}}\sum_{\substack{p_3 \ge p_4 \\ p=p_3+p_4}} \frac{Q_{p1,p2}^p Q_{p3,p4}^{p*} \langle X_{p1} X_{p2} X_{p3}^* X_{p4}^* \rangle}{d}
\end{align}


In order to study the temporal evolution of the fluctuation, the wavenumber Fourier coefficients are required in principle, but the reliable and accurate measurement of wavenumber Fourier coefficients is often challenging. 
Instead, if we assume the Taylor hypothesis, we can exchange between temporal and spatial evolution ($\partial / \partial t = v \partial /\partial z$) in the equation (20) and work with the frequency Fourier coefficients. 
%


\subsubsection{Practical example}

\begin{figure}
    \centering
    \includegraphics[width=0.7\linewidth]{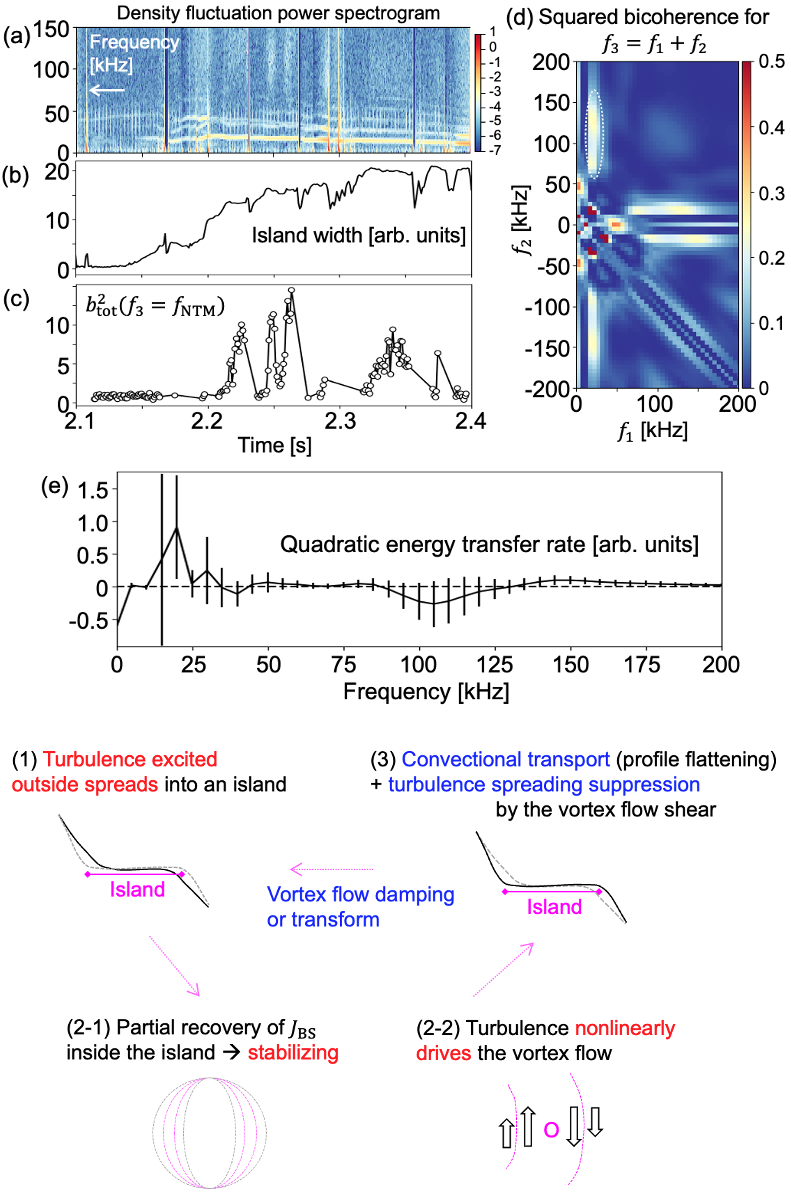}    
    \caption{(Color online) (a) The power spectrogram of density fluctuation near the $m/n=2/1$ neoclassical tearing mode. (b) The $n=1$ mode amplitude from the magnetic probe. (c) The summed total bicoherence with $f_3 = f_\mathrm{NTM}$. (d) The auto-squared bicoherence and (e) the quadratic energy transfer rate of density fluctuation for a strong turbulence period. Reprinted from~\citep{Choi:2021fs}. Copyright 2021 Springer Nature}
    \label{fig:ntm-turb}
\end{figure}

A strong three-wave coupling was observed during the growing phase of the $m/n=2/1$ neoclassical tearing mode (NTM) in DIII-D hybrid H-mode plasmas~\citep{Choi:2021fs}. 
Figure~\ref{fig:ntm-turb}(a) shows the power spectrogram of density fluctuation measured by the BES diagnostics near the NTM during the NTM growing phase.
Figure~\ref{fig:ntm-turb}(b) shows the $n=1$ NTM island width evolution estimated by the magnetic fluctuation analysis.
The $f_\mathrm{NTM}\sim$20 kHz coherent component in figure~\ref{fig:ntm-turb}(a) corresponds to an $n=1$ density perturbation by the NTM, and the broadband (60--150 kHz) fluctuations are observed transiently and repeatedly.
Analysis of the squared bicoherence reveals the strong three-wave coupling between an $n=1$ mode and the broadband fluctuations as shown in figure~\ref{fig:ntm-turb}(d).

The appearance of the strong bicoherence is correlated with the retardation or temporary saturation of the NTM growth.  
The evolution of the summed total bicoherence at $f_3=f_\mathrm{NTM}$ is shown in figure~\ref{fig:ntm-turb}(c).
The strong bicoherence appears in periods when the NTM island width is being temporarily saturated (or slightly decreasing).
This correlation implies that the three-wave coupling might play a role to suppress the NTM growth.\footnote{This three-wave coupling between an $n=1$ mode and the broadband is not a rare observation but it is frequently observed in the NTM growing phase in DIII-D hybrid H-mode plasmas.} 

The quadratic energy transfer rate was estimated using density fluctuation measurements as shown in figure~\ref{fig:ntm-turb}(e), and it indicates that the net quadratic transfer of fluctuation energy for an $n=1$ mode ($\sim$20 kHz) is positive while that for the broadband is largely negative. 
This means the density fluctuation energy flow from the broadband to an $n=1$ mode. 
This result can sound contradicting with the observed correlation between the strong bicoherence and the NTM growth saturation, but rather imply a missing link between the density fluctuation three-wave coupling and the island evolution~\citep{Choi:2021rm}. 
 
One possible scenario to understand these observations was recently suggested~\citep{ChoiAPTWG2023}.
When the island width grows beyond some threshold~\citep{Fitzpatrick:1995ud}, the pressure profile around the island would be modified such that it steepens (flattens) outside (inside) the island.
The steepened pressure profile outside the island can drive plasma turbulence, and the excited turbulence can spread into the island which is linearly stable zone~\citep{Hahm:2004kb}. 
The turbulence spreading and accompanying heat flux~\citep{Choi:2021fs} would modify the pressure profile across the island, relaxing the gradient jump at the island boundary~\citep{Bardoczi:2017gq} (see figure~\ref{fig:ntm-turb}(1)). 
As a result, the bootstrap current inside the island can be partially recovered, which could explain the NTM growth retardation (see figures~\ref{fig:ntm-turb}(2-1) and \ref{fig:ntm-turb}(b)).  
At the same time, the turbulence spreading flux can nonlinearly drive a resonant $n=1$ mode and the vortex flow inside the island~\citep{YoonNF2024} (see figures~\ref{fig:ntm-turb}(2-2), \ref{fig:ntm-turb}(c), and \ref{fig:ntm-turb}(e)).
The convective transport by the vortex flow can again flatten the pressure profile inside the island~\citep{Hornsby:2010fh}, and also the vortex flow shear near the island boundary can suppress the turbulence spreading except the X-point region~\citep{Choi:2017ez, Hahm:2021eb} (see figure~\ref{fig:ntm-turb}(3)).
With the flattened profile, the island can start to grow back, while the vortex flow may be damped or transformed over the long term~\citep{ChoiPRL2022}, eventually returning to its original state.
Although it sounds like a complicated cycle, it could explain both the island width saturation and the density fluctuation energy transfer as results of turbulence spreading into the island and also why this is observed transiently and repeatedly. 

\subsection{Tricoherence to detect four-wave coupling}

The tricoherence is one-order extended from the bicoherence to detect a four-wave coupling~\citep{ChandranIEEE1994}.
The modulogram is a two-dimensional representation of the tricoherence~\citep{KovachTx2022}, devised to detect the spectral coupling among a pump wave $f_p$, an amplitude modulating wave $f_m$, and two sidebands $f_p \pm f_m$.
The auto modulogram can be calculated as follows. 
\begin{equation}
\mathrm{MG} (f_p, f_m) = \langle X^2(f_p) X^*(f_p + f_m) X^*(f_p - f_m) \rangle
\end{equation}

\subsubsection{Practical example}

\begin{figure}
    \centering
    \includegraphics[width=0.5\linewidth]{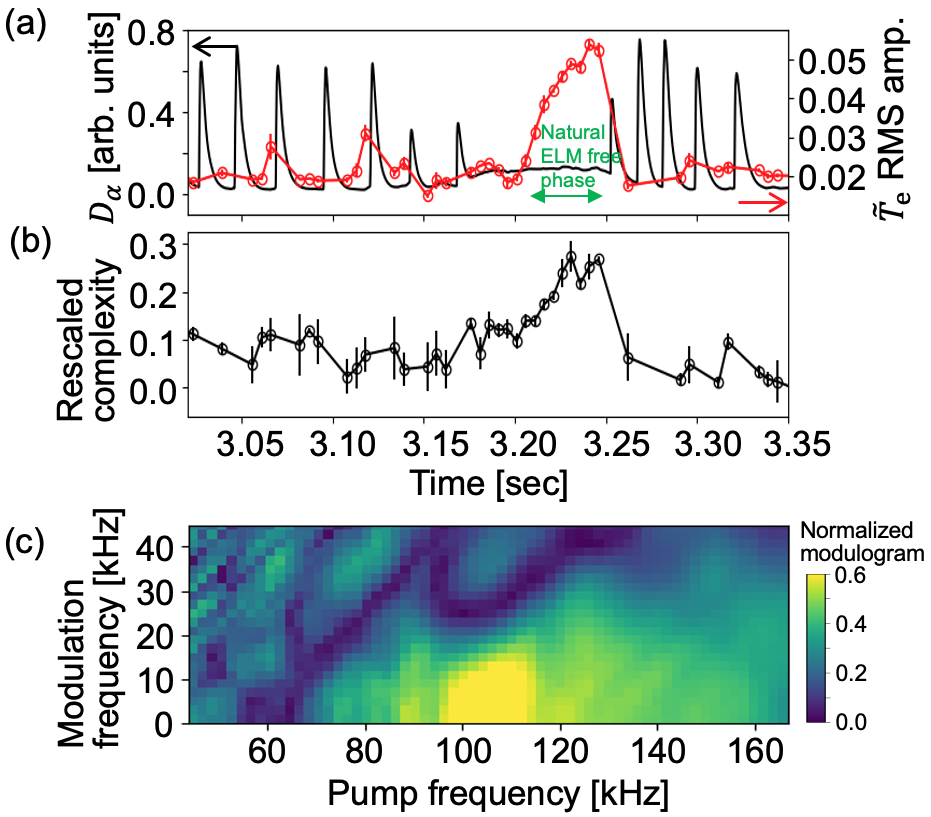}
    \caption{(Color online) (a) The $D_\alpha$ emission (black) and the root mean square (RMS) amplitude of the normalized electron temperature ($T_\mathrm{e}$) fluctuation. (b) The rescaled Complexity of the electron temperature fluctuation. Reprinted from~\citep{ChoiPoP2022}. Copyright 2022 AIP. (c) The modulogram for a strong turbulence period.}
    \label{fig:mod}
\end{figure}

The zonal flow is thought to be responsible for the self-regulation of drift wave turbulence in toroidal plasmas, because one mechanism of the zonal flow generation in toroidal plasmas is via the modulational instability of drift waves~\citep{Chen:2000gf}.
A linearly unstable pump mode is saturated by the nonlinear energy transfer to the sidebands and the zonal flow via the envelop modulation.
Also, according to this theory, the nonlinear oscillations of the drift wave intensity and zonal flows could exhibit a chaotic behavior at the certain parameter regime~\citep{Chen:2000gf}. 

In the KSTAR H-mode plasmas, the growth and saturation of the broadband electron temperature fluctuation near the pedestal top are observed for a short natural ELM-free phase as shown in figure~\ref{fig:mod}(a)~\citep{ChoiPoP2022}. 
The modulogram analysis in figure~\ref{fig:mod}(c) identifies the four-wave couplings among pump modes ($\sim$100~kHz), low frequency modulating modes ($<20$~kHZ), and the sidebands, responsible for the self-regulation of pump modes in the natural ELM-free pedestal.
As consistent with the theoretical observation~\citep{Chen:2000gf}, this strong nonlinearity might also explain the more chaotic nature of the temperature fluctuation in this phase reflected in the increased rescaled Complexity~\citep{ChoiPoP2022} (see below) shown in figure~\ref{fig:mod}(b).

\section{Statistical methods}\label{sec:stat}

\subsection{Gaussian or non-Gaussian}

Spectral methods to detect nonlinear couplings introduced in the previous section would be effective when there exist only a few dominant coupling processes per each triad (or quartet) as in the weak turbulence regime, keeping the (temporarily) coherent phase relationships identifiable.  
In a more complicated system composed of the great number of comparable nonlinear processes as in the strong turbulence regime, it may be difficult to expect one particular nonlinear relationship to be distinguished from others. 
Then, instead of using equations like~\eqref{eq:quad}, the evolution of a physical quantity can be described using a Langevin-type equation where effects of all processes are summed up into a stochastic process. 
\begin{equation}
\phi(t) = \phi_0 + \int_0^t dt^\prime \xi(t^\prime) 
\label{eq:sst}
\end{equation}
where values of $\xi(t)$ are given randomly following some distribution.\footnote{For a white Gaussian noise $\xi_2(t)$, it corresponds to the Brownian motion, or the Wiener process, whose evolution is described by the classical diffusion equation.} 

One may think that the most natural choice for the distribution of $\xi(t)$ is a Gaussian distribution, stemming from the central limit theorem (CLT).  
The CLT states that a sum of $N$ independent and identically distributed random variables, $s_N = \sum_{n=1}^N x_n$, obeys a Gaussian distribution as $N \rightarrow \infty$ when the first and second moments of $x_n$ do not diverge.
It is actually a special case of the limiting ($N \rightarrow \infty$) distribution for the sum $s_N$ when $x_n$ has the finite moments. 
Mathematically, a stable distribution can be the limiting distribution of $s_N$ for some $x_n$, and the Gaussian distribution is one of stable distributions. 
Cauchy and L\'{e}vy distributions are also stable distribution. 
When the distribution of $x_n$ itself has a power-law heavy tail not to have finite first or second moment, the limiting distribution of $s_N$ becomes a L\'{e}vy distribution.
The Gaussian distribution is not the only natural choice for the distribution of $\xi(t)$. 

Analysis of statistical properties of measured turbulence data would provide a hint on the underlying process. 
Calculating the higher order moments (skewness or kurtosis) of the $\phi(t)$ increments, $\Delta \phi (\tau) = \phi (t + \tau) - \phi (t)$, can tell whether the underlying distribution is close to a Gaussian or not.
Skewness (Kurtosis) close to zero (three) corresponds to a Gaussian distribution, and they deviate from those values for a non-Gaussian distribution.

\subsection{Self-similar or intermittent}

The simple stochastic process shown in equation~\eqref{eq:sst} can be expanded to include self-similar processes.
An auto-correlation function of the power-law is often observed in complex systems, $C(\tau) = \tau^{-\gamma}$ satisfying the self-similar condition $C(\tau) = \lambda^{\gamma} C(\lambda \tau)$.  
Especially, when $0 < \gamma < 1$, the system would have an indefinite correlation time and exhibit the long range correlation or anti-correlation (memory effects).   
A power-law kernel is introduced in the stochastic integral to allow memory effects even for a white Gaussian noise $\xi_2(t)$~\citep{Newman:2015jp, Sanchez2018, AlcusonPoP2016}.
\begin{equation}
\phi(t) = \phi_0 + \frac{1}{\Gamma(H- \frac{1}{2} + 1)} \int_0^t dt^\prime (t -t^\prime)^{H - \frac{1}{2}} \xi_2(t^\prime) 
\end{equation}
where $\Gamma(x)$ is the Euler Gamma function and $H$ is the self-similarity exponent or the Hurst exponent~\citep{Sanchez2018}. 
This system corresponds to the fractional Brownian motion (fBm) which evolves either sub-diffusively ($H < 1/2$) or super-diffusively ($H > 1/2$) depending on the value of $H$.
If $H=1/2$, it reduces to the classical Brownian motion evolving diffusively. 

For measured turbulence data resembling the self-similar and Gaussian nature of fBm\footnote{The fractional Brownian motion is non-stationary but it can have the stationary increments, called the fractional Gaussian noise (fGn). The statistical properties of the stationary increments can be more easily compared.}, the Hurst exponent $H$ can be estimated using the conventional method (R/S ratio statistics) to quantify its transport characteristics. 
The expected value of the R/S ratio $E[R/S]$ is defined as follows. 
\begin{equation}
E[R/S](\tau) \equiv \frac{\max\{W(k, \tau); k=1,...,N\}-\min\{W(k,\tau); k=1,...,N\}}{\sqrt{\langle \phi^2 \rangle_\tau - \langle \phi \rangle_\tau^2}}
\end{equation}
where $\tau = t_N$ and $W(k,\tau) =\sum_{j=1}^k (\phi(t_j) - \langle \phi \rangle_\tau)$ is the cumulative series formed by the demeaned $\phi(t_j)$ data. 
The Hurst exponent is given by the scaling exponent $E[R/S](\tau) \sim \tau^{H}$. 


The stochastic process can be further expanded to include intermittent processes. 
Measured turbulence data is often dominated by intermittent events appearing as a heavy tail in the distribution. 
Replacing the Gaussian white noise with an uncorrelated symmetric L\'{e}vy noise $\xi_\alpha (t)$ of a tail exponent $\alpha \in (0,2]$, the stochastic equation can be generalized as~\citep{Newman:2015jp, Sanchez2018, AlcusonPoP2016} 
\begin{equation}
\phi(t) = \phi_0 + \frac{1}{\Gamma(H- \frac{1}{\alpha} + 1)} \int_0^t dt^\prime (t -t^\prime)^{H - \frac{1}{\alpha}} \xi_\alpha(t^\prime) 
\end{equation}
where $0 < H \le \max(1, 1/\alpha)$.
This system corresponds to the fractional L\'{e}vy motion (fLm) which can exhibit both self-similar and intermittent behaviors.\footnote{The fLm is still an mono-fractal model and unable to capture a more irregular intermittent behavior~\citep{Sanchez2018}.} 
When $H = 1/\alpha$ (vanishing the power-law kernel), it reduces to the L\'{e}vy equivalent to the classical Brownian motion, a L\'{e}vy flight. 
It evolves faster (slower) than the L\'{e}vy flight with memory effects if $H > 1/\alpha$ ($H < 1/\alpha$). 

For signals resembling fLm\footnote{The fractional L\'{e}vy motion is non-stationary but it can have the stationary increments, called the fractional L\'{e}vy noise.}, the conventional R/S analysis to estimate the Hurst exponent should be slightly modified due to its diverging variance.   
The denominator is replaced by $1/s$-th power of a moment of order $0<s<\alpha$, i.e. $E[R/S]\equiv (\max\{W(k,\tau)\}-\min\{W(k,\tau)\})/(\langle \phi^s \rangle_\tau - \langle \phi \rangle_\tau^s)^{1/s}$. 
Then, the instantaneous Hurst exponent is calculated $H(\tau):= \frac{\tau}{E[R/S](\tau)} \frac{d E[R/S](\tau)}{d\tau}$.

The degree of intermittency in the series data can be quantified using the so-called intermittence parameter described in the reference~\citep{Carreras:2000ija}.
This value ranges from 0 (mono-fractal) to 1 (multi-fractal). 


\subsubsection{Practical example}

In KSTAR plasmas where the ballistic transport events prevail~~\citep{ChoiNF2019, ChoiPPCF2024}, the self-similarity Hurst exponent and the intermittence parameter are calculated using the measured electron temperature fluctuation in the core region.
The Hurst exponent is found to be larger than 0.5 and the intermittence parameter is about 0.1, meaning that the temperature signal has the long range correlation (see figure~\ref{fig:avalanche} for the power-law frequency spectrum) and it is close to a signal of the mono-fractal system.



\subsection{Stochastic or chaotic}

Whether measured turbulence data is close to signals generated by a stochastic process such as fBm and fGn or chaotic signals generated from deterministic systems having a low dimensional nonlinearity can be distinguished using the Complexity-Entropy analysis~\citep{Rosso:2007vb}.  
Here, Entropy and Complexity are information theoretic terminologies with specific meanings.
Entropy means a measure of missing (unknown) Information of the given probability distribution.
It has the maximum for the equiprobable distribution because one can learn almost nothing from the equiprobable distribution.
Complexity was suggested as the product of Disequilibrium and Entropy (missing Information) to capture the intuitive notion about a complex system~\citep{RuizPLA1995}.
Disequilibrium is a measure of distance from the equiprobable distribution.
The Jensen Shannon divergence is used for Disequilibrium for the Complexity-Entropy analysis~\citep{Rosso:2007vb}.
The idea of Complexity can be understood by two examples~\citep{RuizPLA1995} of simple (not complex) system in physics, i.e. a perfect crystal and an ideal gas.
A perfect crystal, represented by a peaked probability distribution of states, has very small missing Information but large Disequilibrium, and their product Complexity will remain small. 
On the other hand, an ideal gas, represented by an equiprobable distribution of states, has large missing Information but very small Disequilibrium, and their product Complexity also will remain small.
It was shown that signals from various chaotic systems (the logistic map, the skew tent map, Henon's map, etc) and signals from stochastic processes (fBm and fGn) can be separated in the Complexity-Entropy plane~\citep{Rosso:2007vb}. 

The distinction starts with calculating the probability distribution of amplitude orders in partial segments of given time series data, called the Bandt-Pompe (BP) probability distribution~\citep{Pompe:2002jv}. 
Using the BP probability distribution of a given data $P=\{p_j\}_{j=1,...,d!}$ where subscript $i$ represents each amplitude order\footnote{There are $d!$ possible ways of the amplitude ordering for the segment size $d$. The data length should be large enough than $d!$ for the reliable calculation of the BP probability~\citep{Pompe:2002jv}.}, the Jensen Shannon Complexity $C_{\mathrm{JS}}$ and the normalized Shannon Entropy $H$, which are the basis of the Complexity-Entropy analysis~\citep{Rosso:2007vb}, can be calculated.
The Jensen Shannon Complexity ($C_{\mathrm{JS}} = Q H $) is defined as the product of the normalized Shannon Entropy $H=S/S_{\mathrm{max}}$ and the Jensen Shannon divergence $Q$. 
$S=S(P)=-\sum_{j} p_j \ln(p_j)$ is the Shannon Entropy of the given BP probability distribution $P=\{p_j\}$ and $S_{\mathrm{max}}$ is the maximum possible Entropy, i.e. $S_{\mathrm{max}} = \ln(d!)$ with the equiprobable distribution $P_{\mathrm{e}} = \{p_j\} = 1/d!$.
The Jensen Shannon divergence is given as $Q = Q_0 \left\{ S\left(\frac{P+P_{\mathrm{e}}}{2}\right) - \frac{S(P)}{2} - \frac{S(P_{\mathrm{e}})}{2} \right\}$ where $Q_0$ is the normalization constant $Q_0 = -2 / (\frac{d! + 1}{d!} \ln(d! + 1) - 2\ln(2 d!) + \ln(d!))$~\citep{Martin:2006dz}. 
Chaotic signals locate close to the maximum Complexity boundary in the Complexity-Entropy plane, and they move to the locus of points of fBm and fGn signals in proportion to the level of the additive Gaussian noise~\citep{Zhu:2017fo}. 

We suggested the rescaled Complexity as an evaluation metric for the degree of chaos in signals~\citep{ChoiPoP2022}.
It is defined as follows.
\begin{equation}
\hat{C}=\frac{C_{\mathrm{JS}} - C_0}{|C_{\mathrm{bdry}} - C_0|}
\end{equation}
where $C_0(H)$ is the Jensen Shannon Complexity of fBm or fGn and $C_{\mathrm{bdry}}(H)$ is the maximum (if $C_{\mathrm{JS}} > C_0$) or minimum (if $C_{\mathrm{JS}} < C_0$)~\citep{Calbet:2001eq} Jensen Shannon Complexity at the given $H$. 
The rescaled Complexity ($\hat{C}$) ranges from -1 ($C_{\mathrm{JS}} = C_{\mathrm{min}}$) to 1 ($C_{\mathrm{JS}} = C_{\mathrm{max}}$), and the less $\hat{C}$ means the less chaotic or the more stochastic in the relative comparison.

\subsubsection{Practical example}

\begin{figure}
    \centering
    \includegraphics[width=0.5\linewidth]{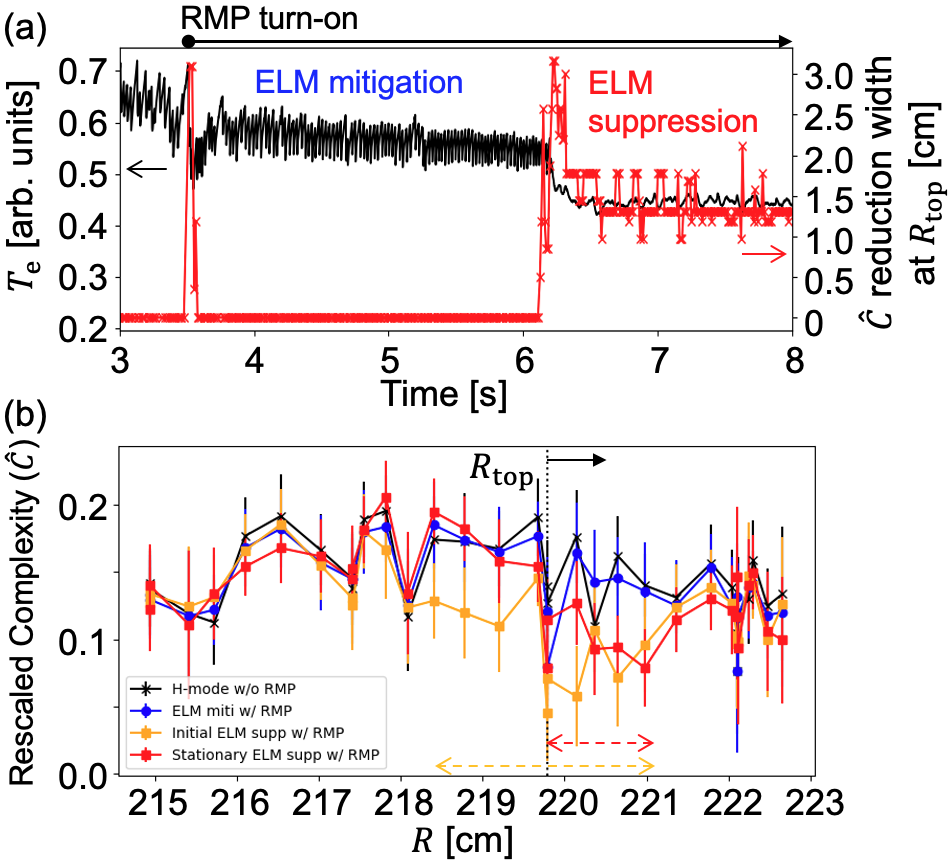}
    \caption{(Color online) (a) The electron temperature and the rescaled Complexity ($\hat{C}$) reduction width near the pedestal top. (b) The rescaled Complexity profiles calculated using two-dimensional temperature fluctuation measurements for different phases. The width of the significant $\hat{C}$ reduction is indicated by the horizontal dashed lines. The pedestal top slightly moves outwards in time by the field penetration. Reprinted from~\citep{ChoiPoP2022}. Copyright 2022 AIP}
    \label{fig:stochastic}
\end{figure}

The rescaled Complexity analysis found that the $T_\mathrm{e}$ fluctuations near the pedestal top become more stochastic with the RMP ELM suppression in KSTAR plasmas~\citep{ChoiPoP2022}.
The significant reduction of the rescaled Complexity ($\hat{C}$) in the $T_\mathrm{e}$ fluctuations is observed for the RMP ELM suppression phase. 
Temperature fluctuation measurements in two-dimensional space are used to obtain the high resolution $\hat{C}$ profiles along the radius on the midplane. 
The width of the significant $\hat{C}$ reduction near the pedestal top is plotted in figure~\ref{fig:stochastic}(a), showing the good correlation with the transition to the ELM suppression phase from the mitigation phase.
Based on the numerical~\citep{Hu:2020ik} and theoretical studies~\citep{Cao:2021, CaoNF2024}, the stochastic fluctuation in the narrow layer ($\sim$1.2 cm in the stationary suppression phase, see figure~\ref{fig:stochastic}(b)) near the pedestal top could be explained as a result of the field penetration and the associated enhancement high-$k$ fluctuations~\citep{ChoiPoP2022}. 
This implies that the high resolution $\hat{C}$ profiles using two-dimensional turbulence data can be used to diagnose the plasma tearing response to the external field~\citep{WilensdorferNP2024}. 


\section{Physics informed neural networks (PINN)}\label{sec:pinn}

\subsection{Missing velocity field prediction using PINN}

An artificial neural network is an attempt by analogy to achieve functions of brain (learning, recollecting, inferring, etc) by mimicking bio-chemical reactions of neurons in brain.
Instead of flows of electric charges, flows of numbers connect an artificial neuron to others. 
Inputs to each neuron are summed with different weights for different connections and determine the output through a nonlinear activation function. 
In the simplest feed-forward form, neurons are placed in layers with one directional hierarchy to construct a network, i.e. outputs of neurons in one layer become inputs to neurons in next layer. 
Eventually, this network composed of multiple layers takes a given input in numbers and return a desired output in numbers through a training. 
Known sets of inputs and outputs are used to update parameters (connection weights and bias) of the network to minimize the difference between the network’s outputs and known outputs, or the loss. 
By increasing the number of neurons and layers, or the dimension of associated parameter hyperspace, the network can learn more complicated features. 

\begin{figure}
\includegraphics[keepaspectratio,width=1.0\textwidth]{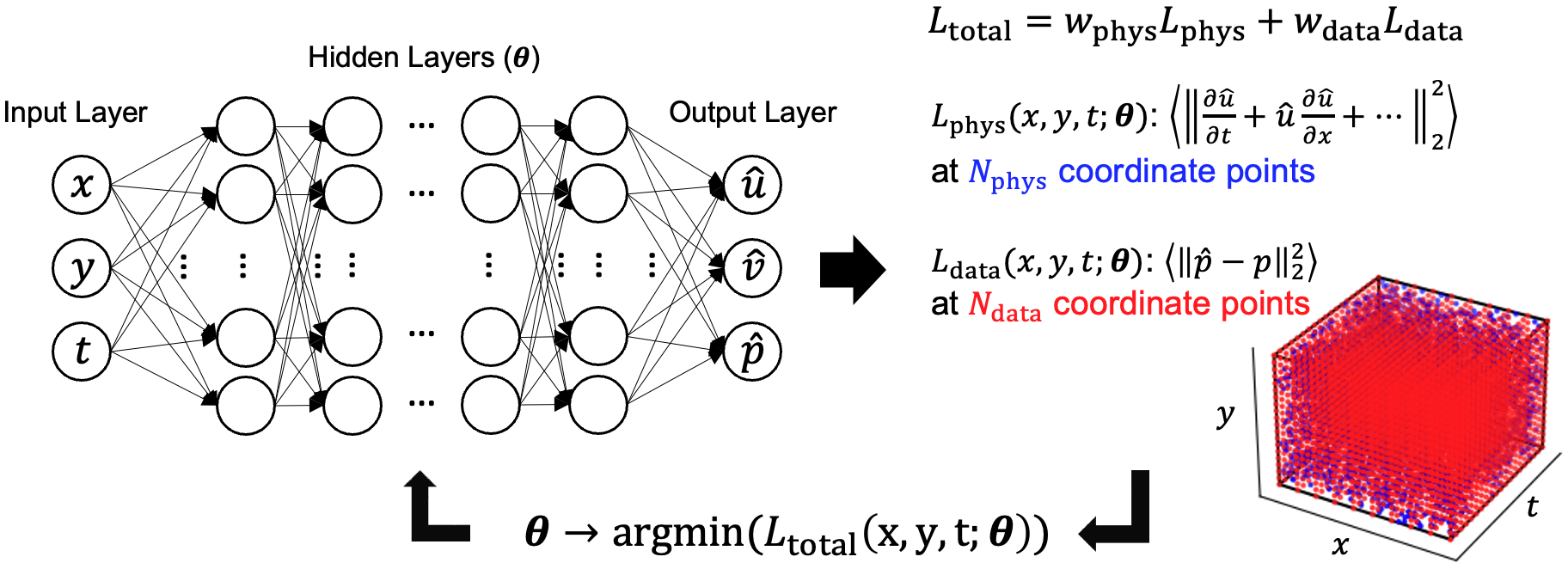}
\centering
\caption{Physics informed neural network for the 2D Navier-Stokes system}
\label{fig:pinn}
\end{figure}

In a physics informed neural network (PINN)~\citep{RaissiS2020}, it takes spatio-temporal coordinates of the system as inputs and returns values of physics variables at the given coordinates as outputs. 
Then, we know that physics variables and their spatial or temporal derivatives satisfy the physics law of the system often expressed in partial differential equations. 
In addition to the data set, this information can be used to train the neural network. 
For example, residuals of the equations for physics variables and their derivatives are added to the loss. 
A neural network trained using both the data and the physics law information is called PINN (see figure~\ref{fig:pinn}).  

PINNs have shown that they can provide sufficiently approximate solution for inverse and ill-posed problems which are difficult to be solved by conventional methods~\citep{SeoSR2024}. 
Also, their performance is more robust against noise in the data than classical neural network~\citep{MathewsPRE2021}. 
These characteristics of PINNs are suited for our purpose, predicting the missing velocity field using the experimental measurements of other physically related fields.  
In the following two examples, we investigate two practical issues in the PINN prediction of a missing velocity field with measured turbulence data, i.e. the noise in the data and the spatial resolution of diagnostic channels, respectively. 

\subsubsection{Effect of the noise in the 2D Navier-Stokes system}

We first consider the 2D $(x, y)$ incompressible Navier-Stokes fluid system where the time $(t)$ evolution of velocity fields $u$ ($x$ component of the fluid velocity) and $v$ ($y$ component of the fluid velocity) and the scalar pressure field $p$ is described by the following equations. 
\begin{align}
0 &= \frac{\partial u}{\partial t} + \left(u \frac{\partial u}{\partial x} + v \frac{\partial u}{\partial y}\right) + \frac{\partial p}{\partial x} - \frac{1}{\mathrm{Re}} \left( \frac{\partial^2 u}{\partial x^2} + \frac{\partial^2 u}{\partial y^2} \right)  \label{eq:ns1} \\
0 &= \frac{\partial v}{\partial t} + \left(u \frac{\partial v}{\partial x} + v \frac{\partial v}{\partial y}\right) + \frac{\partial p}{\partial y} - \frac{1}{\mathrm{Re}} \left( \frac{\partial^2 v}{\partial x^2} + \frac{\partial^2 v}{\partial y^2} \right)  \label{eq:ns2} \\
0 &= \frac{\partial u}{\partial x} + \frac{\partial v}{\partial y}  \label{eq:ns3}
\end{align}
where $\mathrm{Re}$ is the Reynolds number taken as 1 here. 
The first two equations are the 2D Navier-Stokes equations in the absence of body force, and the last equation is the continuity equation for the incompressible fluid. 

This case is known to have the exact 2D solution, called 2D Taylor’s decaying vortices, as follows.
\begin{align}
u(x, y, t) &= -\cos(x) \sin(y) e^{-2t}  \\
v(x,y,t) &= \sin(x) \cos(y) e^{-2t}  \\
p(x,y,t) &= -0.25 (\cos(2x) + \cos(2y)) e^{-4t} 
\end{align}
The computational spatio-temporal domain is bounded as $\pi/2\le x \le 3\pi/2$, $0 \le y \le 2 \pi$, and $0 \le t \le 2$. 
The above solution within this domain is used to generate the hypothetical measurement data sets for training of a PINN. 

We presume that only the pressure measurement is available using the $10 \times 20 = 200$ channels distributed regularly in $(x, y)$ space. 
Total 20 measurements are taken with the uniform time interval to construct the pressure data sets $\{ p_j \}$ at the $(x, y, t)$ coordinates shown as red dots in figure~\ref{fig:pinn}. 
Then, the Gaussian noise of different levels is added to the data to investigate the robustness of a PINN prediction against common additive noise in the actual experimental condition. 

A fully connected feed-forward network with $(x, y, t)$ as inputs and $(u, v, p)$ as outputs is designed using a Python framework, DeepXDE~\citep{LuSiam2021}, to have 7 hidden layers and 64 neurons per each hidden layer: $(\hat{u}, \hat{v}, \hat{p}) = \mathrm{NN}(x, y, t;\pmb{\theta})$ where $\hat{~}$ means the predicted values by the network and $\pmb{\theta}$ indicates all connection weights and bias of the network. 
The network parameters $\pmb{\theta}$ are initialized by the normal Glorot scheme and a tangent hyperbolic function is used for the activation function. 
The total loss of the network is defined as the weighted sum of the pressure data loss $L_\mathrm{data} =\frac{1}{N_\mathrm{data}}\sum_j \| \hat{p}(x_j, y_j, t_j) - p(x_j, y_j, t_j)\|_2^2$ and the physics loss $L_\mathrm{phys} = \frac{1}{N_\mathrm{phys}} \sum_j \| f(x_j, y_j, t_j; \hat{u}, \hat{v}, \hat{p}; \frac{\partial \hat{u}}{\partial x}, \frac{\partial \hat{u}}{\partial y}, \frac{\partial \hat{u}}{\partial t}, \cdots )\|_2^2$.
$f$s are the residuals of equations~\eqref{eq:ns1}--\eqref{eq:ns3} at the randomly distributed $(x, y, t)$ coordinates shown as blue dots in figure~\ref{fig:pinn}. 
$N_\mathrm{data}$ and $N_\mathrm{phys}$ mean the number of coordinate points used to calculate each loss. 
Automatic differentiation implemented in TensorFlow~\citep{Abadi2016} is used within the DeepXDE framework~\citep{LuSiam2021} to calculate derivatives of outputs with respect to inputs. 
Weighting the physics loss higher (10--30 times) than the data loss ($w_\mathrm{phys} > w_\mathrm{data}$) results in the better training with the missing data. 
The PINN is trained by successive optimization of the network parameters to reduce the total loss using Adam~\citep{Tieleman2012} (learning rate is set as 1e-3) and L-BFGS~\citep{Bonnans2006} algorithms. 
The NVIDIA CUDA acceleration facilitated the training on the KAIROS-GPU server in KFE. 

\begin{figure}
\includegraphics[keepaspectratio,width=0.75\textwidth]{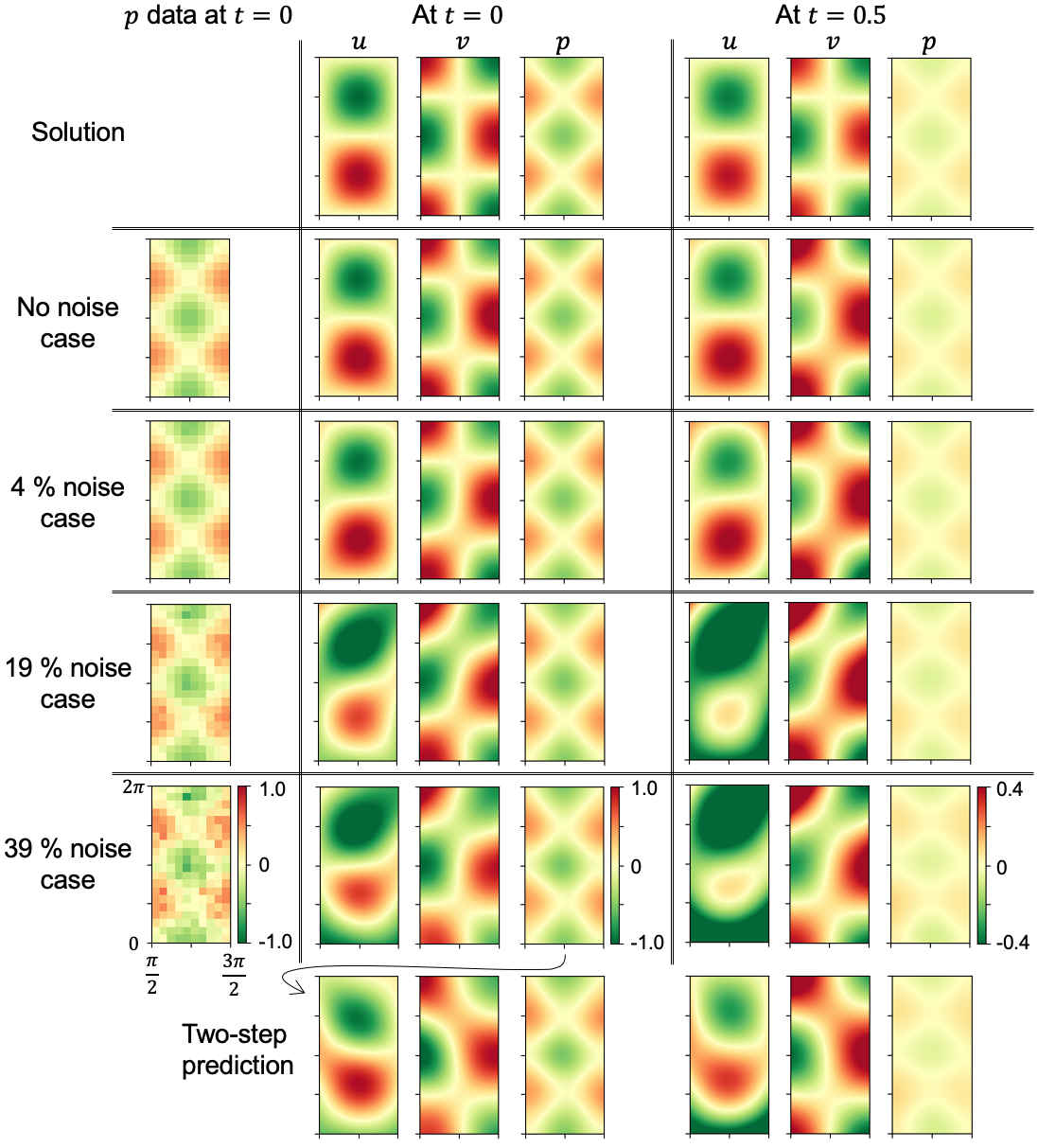}
\centering
\caption{The PINN prediction of the 2D Navier-Stokes system}
\label{fig:ns}
\end{figure}

Results of the PINN prediction of velocity $u$, $v$ and pressure $p$ fields of the 2D Navier-Stokes system are summarized in figure~\ref{fig:ns}. 
The solutions at two different times $t=0$ and $t=0.5$ are plotted at the top, showing rapidly decaying vortices (the narrower color scales were used at the later time). 
Each PINN, $(\hat{u}, \hat{v}, \hat{p}) = \mathrm{NN}(x,y,t;\pmb{\theta})$, is trained with the pressure data sets $\{ p_j \}$ at the same $(x, y, t)$ coordinates including different amount of noise. 
Examples of the pressure data at $t=0$, with the Gaussian noise of different standard deviations, are shown in the first column. 
The PINN predictions of $u$, $v$, and $p$ at $t=0$ and $t=0.5$ are shown for different levels of the noise in the corresponding columns and rows (4~\% means that the standard deviation of the noise is 4~\% of the standard deviation of the pure data). 

Figure~\ref{fig:ns} shows that the PINN can predict the velocity fields with only the pressure data provided thanks to the additional physics information used in the training. 
Obviously, the PINN prediction shows the better agreement with the lower noise level, but it is notable that the results with the 39~\% noise level still capture the characteristic vortex structure. 
As time flows and the vortex decays, the deviation becomes more evident with the larger noise. 

The pressure prediction seems to be more robust against the noise than the velocity fields prediction. 
This implies that we can design a two-step noise-robust PINN prediction of the velocity fields: The first PINN aims to refine the pressure data, and the second PINN predicts the velocity fields with the refined pressure data generated by the first PINN (with the arbitrarily resolution in the $(x, y, t)$ space). 
The result shown at the bottom of figure~\ref{fig:ns} corresponds to the two-step PINN prediction using the pressure prediction of the 39~\% noise case as the training data of the second PINN. 
To prevent the second PINN from converging to the same minimum loss point in the parameter hyperspace as the first PINN, the second PINN should be configured slightly differently (it can have different activation function such as the swish function or additional hidden layer). 
Compared to the result of the 39~\% noise case, the two-step prediction shows the improvement, meaning a more noise-robust PINN prediction of velocity fields with the pressure data alone in the 2D Navier-Stokes system. 

\subsubsection{Effect of the channel resolution in the 2D Hasegawa-Wakatani system}

Next, we consider the 2D Hasegawa-Wakatani system where the time $(t)$ evolution of the density ($n$), the electrostatic potential ($\phi$), and the vorticity ($\Omega$) fields is described by the following equations.
\begin{align}
0 &= \frac{\partial n}{\partial t} - c_1 (\phi - n) + [\phi, n] + \kappa_n \frac{\partial \phi}{\partial y} + \nu \nabla^{4} n   \\
0 &= \frac{\partial \Omega}{\partial t} - c_1 (\phi - n) + [\phi, \Omega] + \nu \nabla^{4}\Omega  \\
0 &= \Omega - \nabla^2 \phi  
\end{align}
where $[a,b] = \frac{\partial a}{\partial x} \frac{\partial b}{\partial y} - \frac{\partial b}{\partial x} \frac{\partial a}{\partial y}$, $c_1$ is the adiabatic coefficient taken as 1, $\kappa_n$ is the density gradient scale length taken as 1, and $\nu$ is the diffusion coefficient taken as 5e-6. 
Although this set of equations may not be sufficiently sophisticated, it contains fundamental dynamics and complexity of the drift wave turbulence in fusion plasmas. 
It can be a minimal test bed for the PINN prediction of turbulent velocity field (the $E\times B$ velocity $\bold{v}_{E\times B}=\frac{1}{B^2} \bold{B} \times \nabla \phi$ where $\bold{B}$ is the magnetic field vector) with limited density turbulence measurements.     

The equations are numerically solved using a verified Python solver called HW2D~\citep{GreifJOSS2023}. 
The simulation is run sufficiently long ($t \le 500$: the nonlinear saturation phase is reached around $t\sim 150$ and the calculation time step $\delta t = 0.025$) over the $0 \le x \le 41.902$ and $0 \le y \le 41.902$ space with 512 x 512 grid points ($\delta x = \delta y = 0.082$), but the domain of interest for the PINN training is restricted as $0 \le x \le 4.1$, $0 \le y \le 2.05$, and $309 \le t \le 321$ due to the limited computational memory. 

We presume that only the density measurement is available from the channels distributed regularly in $(x, y)$ space. 
The number of channels is varied to investigate the effect of the spatial resolution of diagnostic channels for the reasonable PINN prediction with the fourth-order derivatives in the equations. 
Measurements are taken with the uniform time interval to construct the density data sets $\{ n_j \}$ at the $(x, y, t)$ coordinates. 

The neural network of the same architecture as in the previous example is used: $(\hat{n}, \hat{\phi}, \hat{\Omega}) = \mathrm{NN}(x,y,t;\pmb{\theta})$. 
It is left for future work to explore the possibility of the improvement by changing the network architecture and relevant hyperparameters.  

\begin{figure}
\includegraphics[keepaspectratio,width=1.0\textwidth]{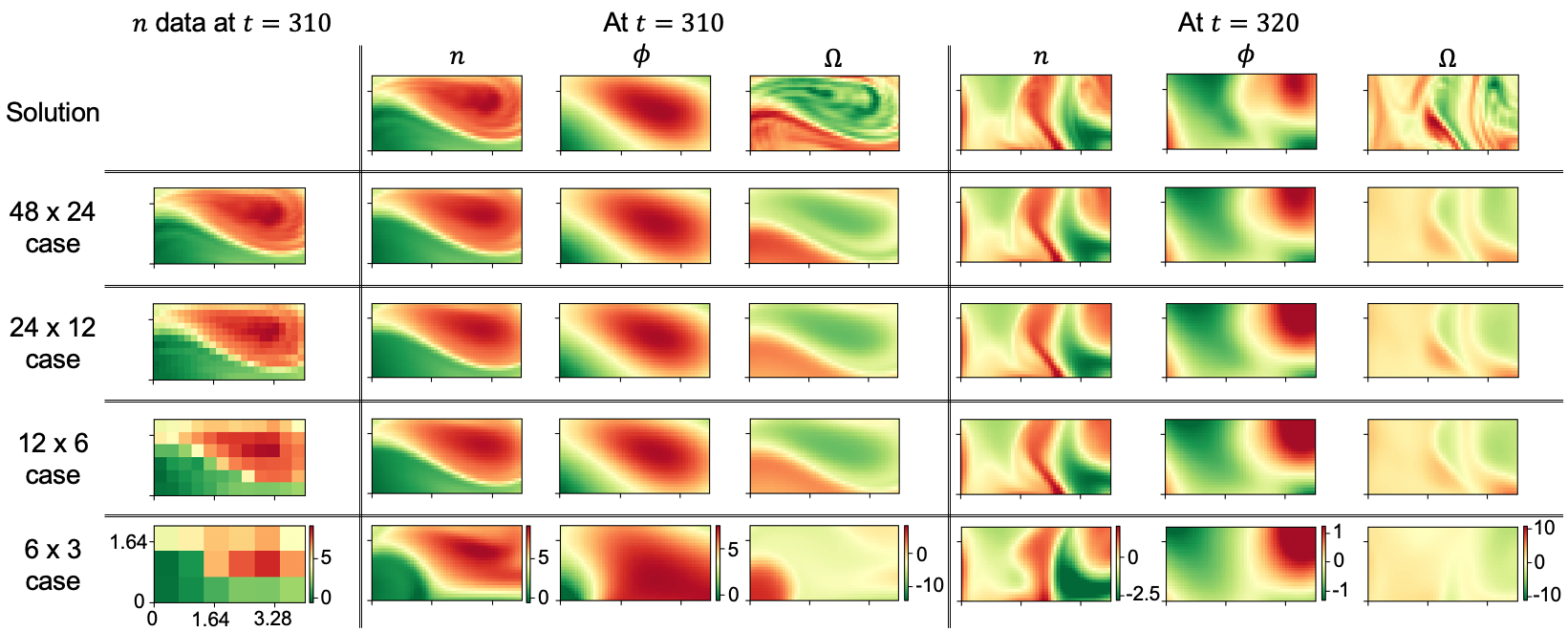}
\centering
\caption{The PINN prediction of the 2D Hasegawa-Wakatani system}
\label{fig:hw}
\end{figure}

Results of the PINN prediction of the density $n$, the electrostatic potential $\phi$ (the $E\times B$ velocity $\bold{v}_{E\times B}=\frac{1}{B^2} \bold{B} \times \nabla \phi$) and the vorticity $\Omega$ fields of the 2D Hasegawa-Wakatani system are shown in figure~~\ref{fig:hw}. 
The solutions at $t=310$ and $t=320$ are plotted at the top, compared with the predictions by the PINNs trained with the density data sets of different spatial resolutions. 
Each PINN, $(\hat{n}, \hat{\phi}, \hat{\Omega}) = \mathrm{NN}(x,y,t;\pmb{\theta})$, is trained with the density data sets $\{ n_j \}$ obtained by different number of channels in the $(x,y)$ space. 
Due to the limited computational memory, the narrower time domain was chosen for the larger channel number case, but the temporal resolution of measurements was kept constant as 5 measurements per $\Delta t =1$. 
Examples of the density data at $t=310$ are shown in the first column for different channel number cases. 
The PINN predictions of $n$, $\phi$, and $\Omega$ at two times are shown for different channel number cases in the corresponding columns and rows. 
The 48 x 24 case means 48 (24) channels are uniformly distributed along $x$ ($y$) direction and separated by $\Delta x = \Delta y = 0.089$. 

With given only the density measurements, the PINN could capture the key characteristics of the potential field down to the 12 x 6 case. 
Some fine structure in the density field could be reproduced in the 48 x 24 case, but as the number of channels decreases it is lost in the prediction. 
On the other hand, the fine structure in the vorticity field could not be reproduced at all. 
Changing the network architecture or renormalizing fields to be on the same scale might be helpful to improve the prediction of the vorticity field.

\subsubsection{Point spread function effects}

Besides the noise and the spatial resolution, the point spread function of the diagnostic channel~\citep{GhimRSI2010, Fox:2017dr} is another crucial factor that must be taken into account when applying the PINN velocity predictor in practice. 
Turbulence diagnostics in fusion plasmas such as BES and ECEI utilize optics and analogue/digital filters to receive a signal whose power depends on plasma fields such as density and temperature in a region of finite volume. 
In other words, the measured signal of the diagnostic channel contains the integrated information over the measurement region. 
The point spread function of a channel relates local values of plasma fields in the region with the measured signal by the channel. 
It should be narrow enough not to average out the spatial variation in the structure of interest, but also large enough for the measured signal power not to be buried in noise. 
Due to effects of the point spread function, the measured signal of a diagnostic channel does not exactly reflect local values of plasma fields which would be ideal for the training of the PINN velocity predictor.

One may consider effects of the point spread function as noise in the data and anticipate that the PINN can predict a velocity field directly using the measured data by diagnostics. 
Or, one can consider deduction of the plasma fields from the measured data as another inverse problem that can be solved by an artificial neural network. 
All data sets (the prescribed density and temperature field and the corresponding synthetic data of diagnostics) for the training of the network can be freely generated using forward modeling tools~\citep{syndia}. 
Then, a two-step prediction of the velocity field would be possible: from the measured turbulence data by diagnostics to the density or temperature field and from the density or temperature field to the velocity field using the PINN as shown above. 

\subsection{Model validation using measured spatio-temporal dynamics of turbulence}

The PINN can inversely identify a solution satisfying physics model equations of a turbulence system with the given measurements, unless the noise is too strong or the spatial resolution of the measurements is insufficient. 
This implies a way to test model equations with measured turbulence data. 
Due to the chaotic behavior of turbulence, it was not meaningful to directly compare the observed spatio-temporal dynamics of turbulent field in experiments with that obtained in the forward calculation of model equations. 
It is almost impossible to match all the initial and boundary conditions between the experiment and the simulation, and so the reproducible statistical quantities in the stationary state are compared at best~\citep{ChoiPPCF2024}. 
However, the initial and boundary condition determine which stationary state a system reaches~\citep{Dif-PradalierCP2022}. 
By comparing the minimum physics losses between PINNs constrained with the turbulence synthetic data from the model and the experimentally measured data, it could be decided whether a model is appropriate to describe the evolution of measured plasma field in the experiment or not.

\section{Summary}

Methods for leveraging fluctuation measurements in fusion experiments are introduced with some practical examples.
The spectral methods are useful to identify different modes and investigate a dominant nonlinear coupling among them. 
When the degree of nonlinearity is too large to be analyzed using the spectral methods, it can be viewed as a stochastic system and the statistical methods can be used to understand the characteristics and behavior of the system. 
On the other hand, recent developments of the physics informed neural network allow leveraging two-dimensional fluctuation measurements in novel ways, predicting a missing field or validating turbulence models. 
Given a validated physics model, the PINN velocity predictor would accurately reconstruct a high-resolution velocity field from other measured turbulence data, provided that the data have a sufficient signal-to-noise ratio and spatial resolution. 
These method would contribute to a better understanding of plasma turbulence transport in fusion experiments.  

\section*{Code availability}
The Python code package named as ``fluctana''~\citep{fluctana} has been developed to provide an easy access and analysis of fluctuation data from fusion experiments. 
This includes most of spectral and statistical methods introduced in this paper. 
The codes and simple tutorials are available in the GitHub repository~\url{https://github.com/minjunJchoi/fluctana}. 

\begin{acknowledgements}
The author (M.J.C.) is grateful to Prof. H.K. Park, Prof. G.S. Yun, Dr. W. Lee, Dr. S. Zoletnik, and Prof. Y.-c Ghim, from whom he learned the fundamentals of plasma diagnostics and data analysis.
He also thanks Prof. T.S. Hahm, Prof. P.H. Diamond, Dr. J.-M. Kwon, Dr. H. Jhang, Dr. Juhyung Kim, Dr. K. Ida, Prof. Y. Kishimoto, Prof. E.-J. Kim and Dr. G. Dif-Pradalier for their valuable discussions on the scientific results introduced above. 
The author (M.J.C.) further appreciates the supports from Mr. J. Park and Dr. G. Jo for the KAIROS-GPU.
This research was supported by R\&D programs of ``High Performance Tokamak Plasma Research \& Development (code No. EN2501)'' and ``High Performance Fusion Simulation R\&D (EN2541)'' through Korea Institute of Fusion Energy (KFE) funded by the Government funds, Republic of Korea.
Computing resources for the neural network training were provided on the KFE computer, KAIROS-GPU, funded by the Ministry of Science and ICT of the Republic of Korea (EN2541).
\end{acknowledgements}

\bibliographystyle{spbasic} 

\end{document}